\documentclass[a4paper,11pt]{article}
\usepackage{jinstpub} 
\usepackage{lineno}
\usepackage{url}
\usepackage{hyperref}
\usepackage{doi}

\title{\boldmath Improved liquid argon ionization model and its impact on the DarkSide low-mass WIMP search programme}

\collaboration[c]{on behalf of DarkSide-50 and DarkSide-20k collaborations}

\author{D. Franco}
\affiliation{APC, Universit\'e  Paris Cit\'e, CNRS, Astroparticule et Cosmologie, Paris F-75013, France}

\emailAdd{davide.franco@apc.in2p3.fr}

\abstract{DarkSide-50 achieved leading WIMP limits down to 1.2 GeV/$c^{2}$ with an ionization-only analysis, despite its small active mass of 46 kg compared to multi-ton noble-liquid detectors. Accurate modelling of the nuclear-recoil ionization yield ($Q_y$) is central to interpreting such searches. A new global fit combining nuclear-recoil response measurements from DarkSide-50, ARIS, SCENE, and the recent ReD experiment constrains $Q_y$ between 0.4 and 150 keV within the Thomas–Imel box framework. The dependence on screening potentials is addressed through a Bayesian model comparison, which rejects the ZBL and Molière functions and strongly favours the Lenz–Jensen one. The resulting model improves DarkSide-50 sensitivity below 3 GeV/$c^{2}$ and refines the DarkSide-20k sensitivity accordingly.
}


\begin{document}
\maketitle
\flushbottom

\section{Ionization response of liquid argon to low-energy nuclear recoils}

The search for dark-matter particles through their elastic scattering off atomic nuclei is the primary goal of the DarkSide programme, which employs dual-phase liquid-argon time projection chambers (LAr TPCs). These detectors measure both the primary scintillation (S1) and ionization (S2) signals, providing powerful discrimination between electronic and nuclear recoils. The DarkSide-50 experiment operated between 2012 and 2019 with a 46~kg active mass of underground argon at LNGS~\cite{DarkSide:2018kuk}. Thanks to the ionization-only analysis, which allows to lower down the threshold to the sub-keV range, DarkSide-50 has set world-leading limits on WIMP--nucleon interactions between 1.2 and 3.5~GeV/$c^{2}$~\cite{DarkSide-50:2022qzh}.  

In this low-mass regime, the interpretation of the data critically depends on the accurate modelling of the ionization yield $Q_{y}$, defined as the number of ionization electrons produced per unit recoil energy. To avoid confusion between the model and the data, we use $f_q(E)$ to indicate the model prediction for the ionization yield, while $Q_y$ denotes the experimentally measured quantity. In the DarkSide-50 analysis, $Q_{y}$ was described within the Thomas--Imel box model and constrained by neutron calibrations with AmBe and AmC sources, together with external datasets from ARIS and SCENE experiments~\cite{DarkSide:2021bnz}. 
We note that the DarkSide-50 in-situ AmC neutron calibration already provides direct experimental constraints on the nuclear-recoil ionization response down to $\sim$3 extracted electrons, corresponding to recoil energies of about 0.4~keV, and was shown to strongly limit any significant suppression of $Q_y$ at low energy. This point, together with the associated model uncertainties at the sub-keV scale, was extensively discussed in Ref.~\cite{DarkSide:2021bnz}. However, the model predictions at such low energies remained dependent on the assumed nuclear stopping power, $s_{n}$, as follows

\begin{equation}
f_{q}(E) \;=\; 
\frac{F}{E\,C_{\rm box}}\,
\ln\!\left[
1 \;+\;
\epsilon\, \beta\, \frac{C_{\rm box}}{F}\,
\frac{s_{e}(\epsilon)}
     {s_{e}(\epsilon) + s_{n}(\epsilon)}
\right],
\label{eq:fq}
\end{equation}
\noindent with $\epsilon$  ($\sim$0.0135 $E/\mathrm{keV}$ in LAr) the dimensionless reduced energy, $s_{e}$ the electronic stopping power, $F$ the electric field,  $C_{\rm box}$ the box constant proportional to the effective size of the ionization cloud,  and $\beta$ a proportionality factor relating energy deposition and initial ion–electron pair density. 

The nuclear stopping power depends on  the screening functions (SFs), which determine the energy transferred during atomic collisions. A previous DarkSide analysis~\cite{DarkSide:2021bnz} explored whether existing measurements could already distinguish among competing screening potentials. That study combined low-energy recoil data from DarkSide-50 (down to $\sim$0.4 keV) with the monoenergetic recoil measurements provided by the ARIS~\cite{Agnes:2018mvl} and SCENE~\cite{Cao:2014gns} experiments, with  energies as low as 7 keV.  However, the combined dataset did not offer enough sensitivity to favor any of the commonly used models, namely the ZBL (or Ziegler \textit{et al.})~\cite{Ziegler:2010bzy}, the Molière~\cite{Moliere:1947zza}, and the Lenz–Jensen functions~\cite{Lenz,Jensen}. 
The ZBL model is semi-empirical and widely used in ion-implantation calculations, whereas both the Molière~\cite{Moliere:1947zza} and Lenz–Jensen~\cite{Lenz,Jensen} functions originate from Thomas–Fermi theory. Molière introduced a three-exponential approximation to the Thomas–Fermi potential, while Lenz and Jensen proposed a simpler analytic alternative.

\begin{figure}[htbp]
\centering
\includegraphics[width=.7\textwidth]{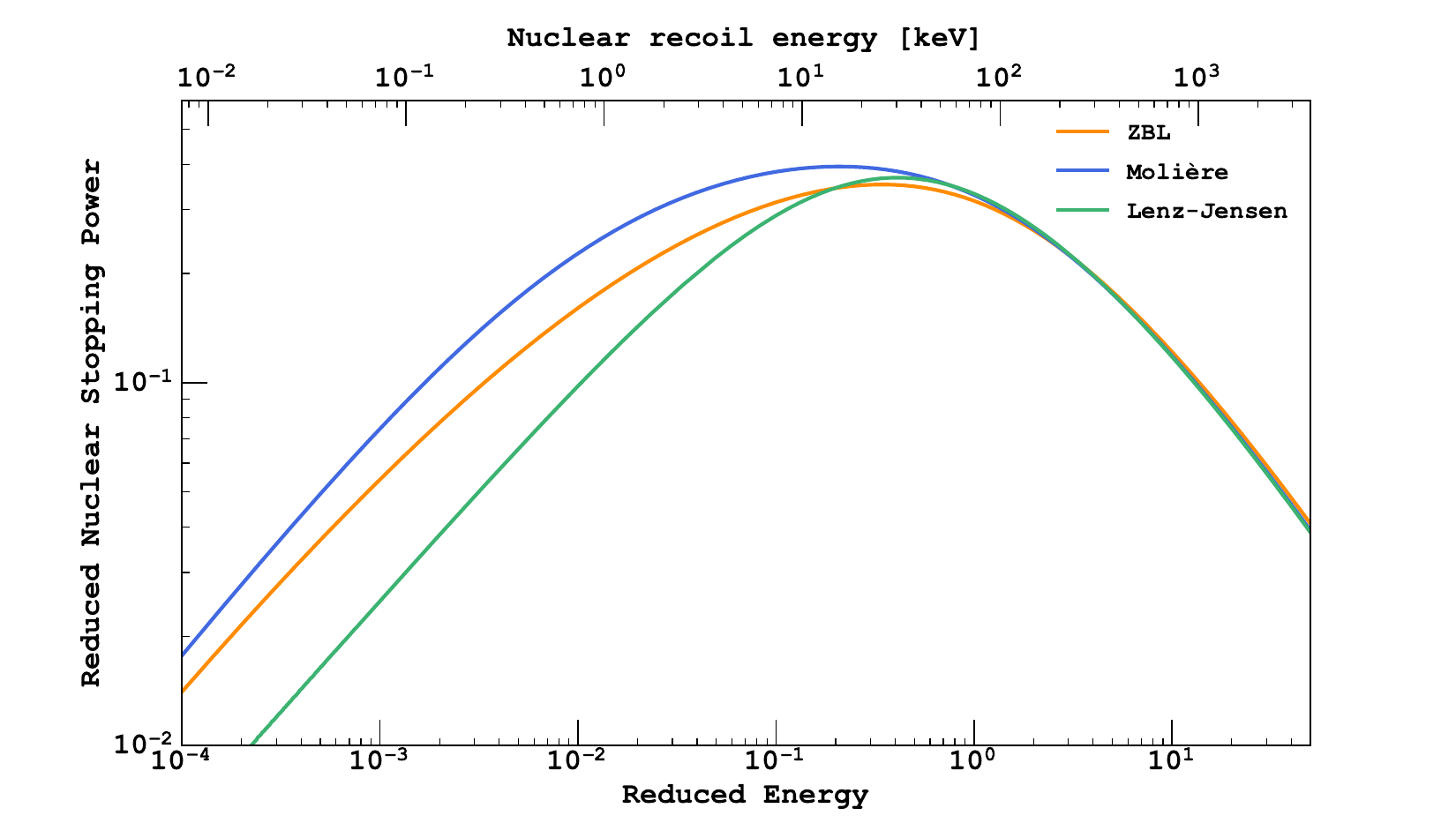}
\caption{Reduced nuclear stopping power as a function of reduced energy (bottom axis) and nuclear recoil energy in liquid argon (top axis), for the three screening potentials adopted in this work: ZBL, the Molière model implemented with the Wilson parametrization, and the Lenz–Jensen function. }
\label{fig:screening}
\end{figure}

In earlier DarkSide-50 work~\cite{DarkSide:2021bnz} and in the analysis of Bezrukov et al.~\cite{Bezrukov:2010qa}, the Molière model was implemented using the approximation published by Winterbon~\cite{Winterbon1972}, based on numerical coefficients that were later found to be incorrectly reproduced in the widely used Sigmund compilation~\cite{Sigmund2014}.
Moreover, even when correctly tabulated, the Winterbon approximation is known to deviate noticeably from the true Molière stopping power.
To overcome these issues, we adopt the parametrization proposed by Wilson et al.~\cite{Wilson1977}, which reproduces the Molière screening behaviour at the percent level and enables a more reliable comparison among the different models. The comparison between the screening functions is shown in Fig.~\ref{fig:screening}. 

\begin{figure}[htbp]
\centering
\includegraphics[width=.7\textwidth]{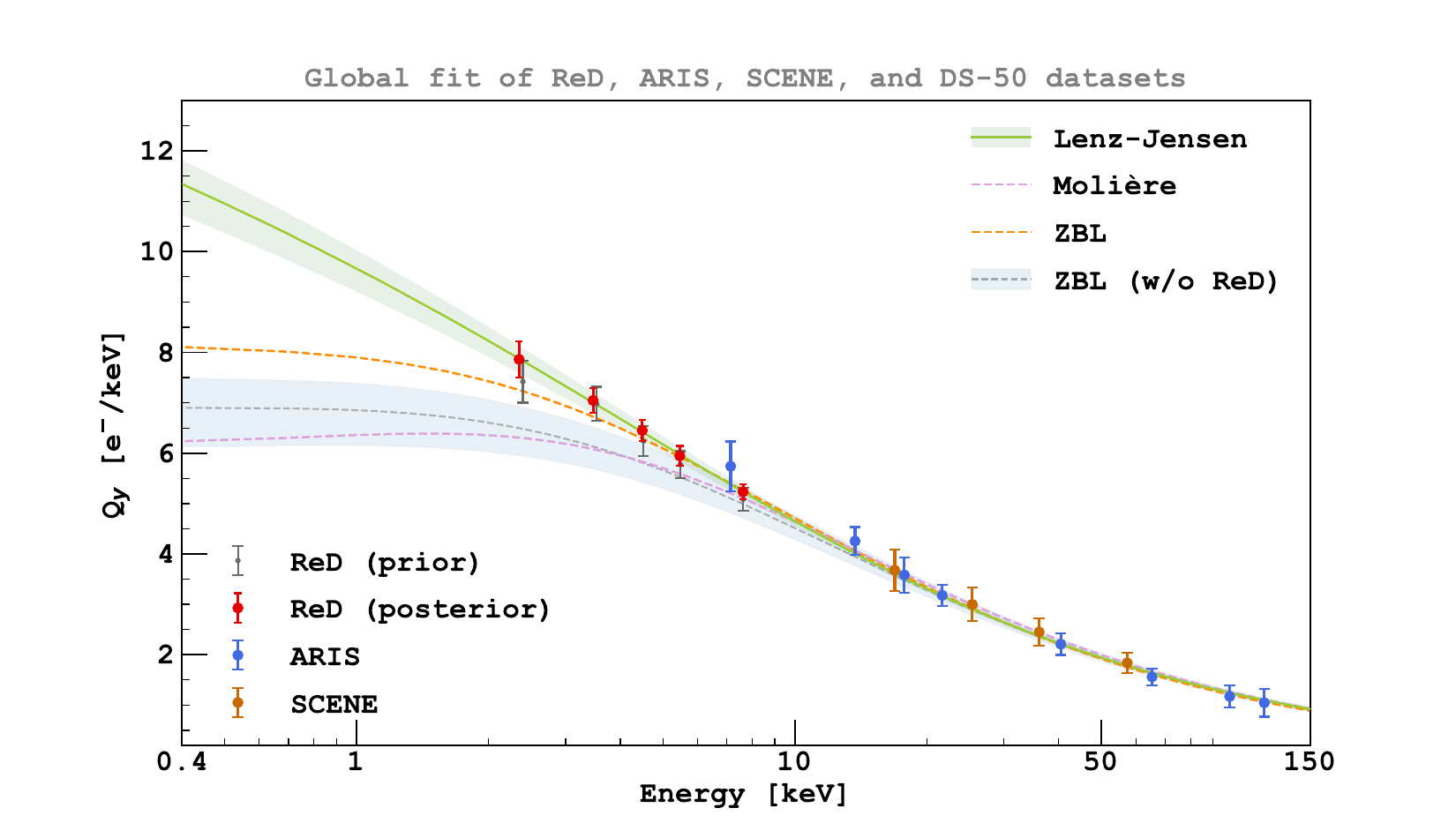}
\caption{Simultaneous fit to the ReD, ARIS, SCENE, and DarkSide-50 datasets assuming the Lenz–Jensen screening (green). ReD points are shown with prior (gray) and posterior (red) uncertainties. The previous ZBL-based model from ref.~\cite{DarkSide:2021bnz} is shown in gray, and global fits using ZBL (orange) and Molière (purple) screening functions are overlaid for comparison~\cite{DarkSide-50:2025lns}.}
\label{fig:qy}
\end{figure}


A key advancement presented is the inclusion of new measurements from the ReD experiment~\cite{DarkSide-50:2025umf}, a small dual-phase LAr TPC designed to provide event-by-event nuclear recoil energies through neutron time-of-flight from a $^{252}$Cf source. The resulting $Q_y$ data span the 2--10~keV range with uncertainties at the few-percent level, precisely where the screening potential has the largest impact on the ionization response. We incorporate these measurements into an updated global fit of the ionization model, performed simultaneously with DarkSide-50 AmC data and the monoenergetic recoil measurements from ARIS and SCENE.

The fit follows the framework developed in ref.~\cite{DarkSide:2021bnz}, varying the two model parameters $C_{\mathrm{box}}$ and $\beta$ while fixing the drift field to 200~V/cm from eq.~\ref{eq:fq}. For ReD, the nuisance parameters associated with the ionization-electron amplification gain in the gas phase  and with a possible vertical offset of the TPC relative to the cryostat center  are assigned Gaussian priors and are marginalised over when constructing the global $\chi^2$~\cite{DarkSide-50:2025lns}. The inclusion of ReD significantly improves the sensitivity of the fit to the choice of screening potential, allowing the model to be tested in the recoil-energy region most relevant for light WIMPs. 

The combined dataset  favours the Lenz–Jensen screening function over both ZBL and Molière, as can be already inferred from the fit results shown in Fig.~\ref{fig:qy}. This conclusion is supported quantitatively by the Bayes factors (BFs) computed from the marginal likelihoods of the three non-nested models. The data prefer Lenz--Jensen over ZBL with  $\log_{10}\mathrm{BF}=3.8$, corresponding to odds of nearly $10^{4}\!:\!1$, and over Molière with $\log_{10}\mathrm{BF}=7.2$, i.e.\ a preference exceeding $10^{7}\!:\!1$.   According to standard model-selection criteria, these values constitute decisive evidence.  

Together, these results establish Lenz--Jensen as the only screening model capable of consistently reproducing the full ensemble of ReD, ARIS, SCENE, and DarkSide-50 measurements.

\section{Impact on DarkSide-50 limits and DarkSide-20k sensitivity}

The updated ionization response based on the Lenz--Jensen screening function has a direct impact on the interpretation of low-energy nuclear recoils and therefore on the sensitivity of liquid-argon experiments to low-mass WIMPs. We re-evaluated the DarkSide-50 exclusion limits by replacing the previously adopted ZBL-based   with the new best-fit Lenz-Jensen-based model. Following the same statistical framework of ref.~\cite{DarkSide-50:2022qzh}, the limits are computed through a binned profile-likelihood analysis that includes all detector-related systematic uncertainties, the measured single-electron response, and the established background models.

\begin{figure}[htbp]
\centering
\includegraphics[width=.45\textwidth]{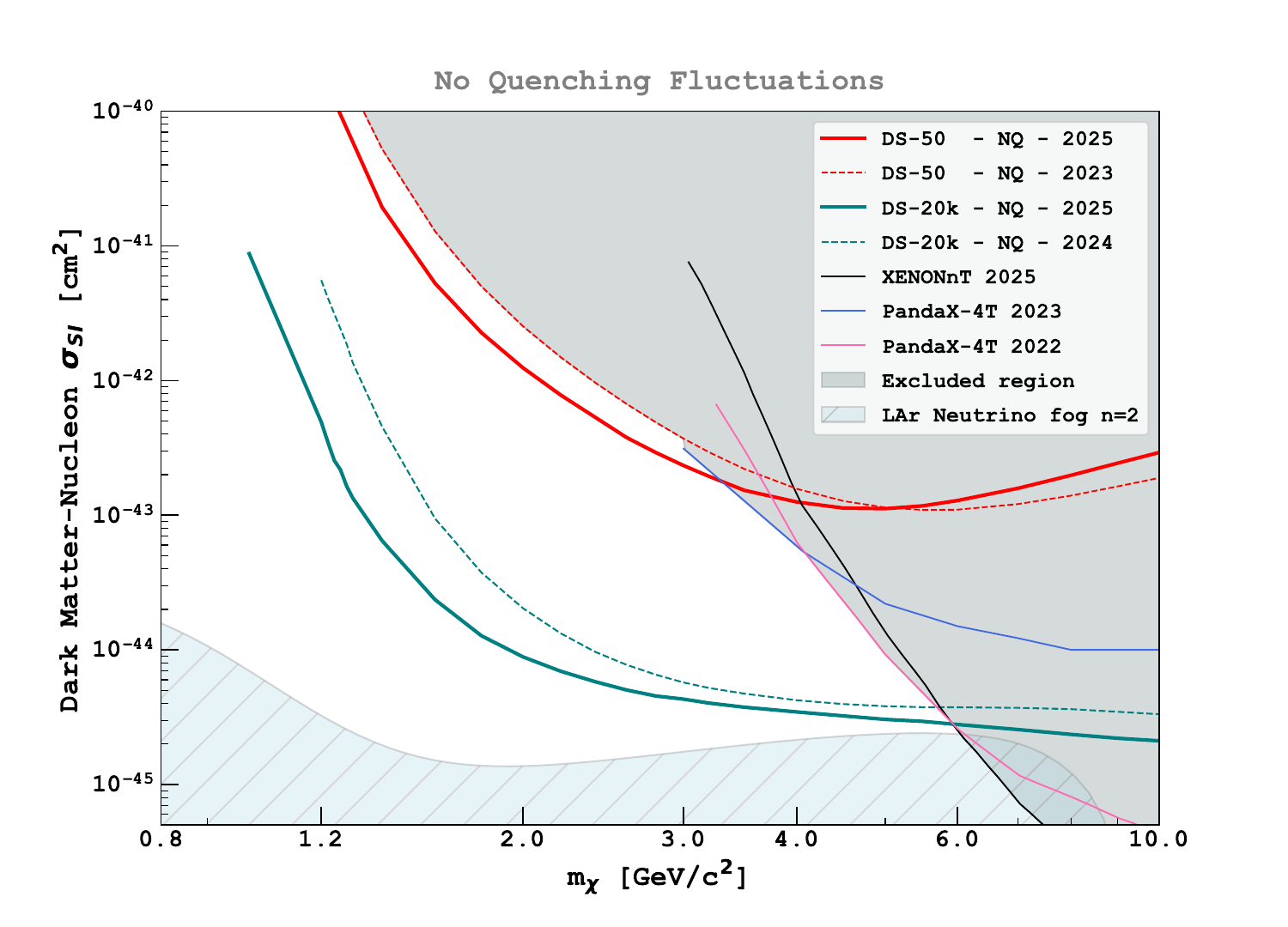}
\includegraphics[width=.45\textwidth]{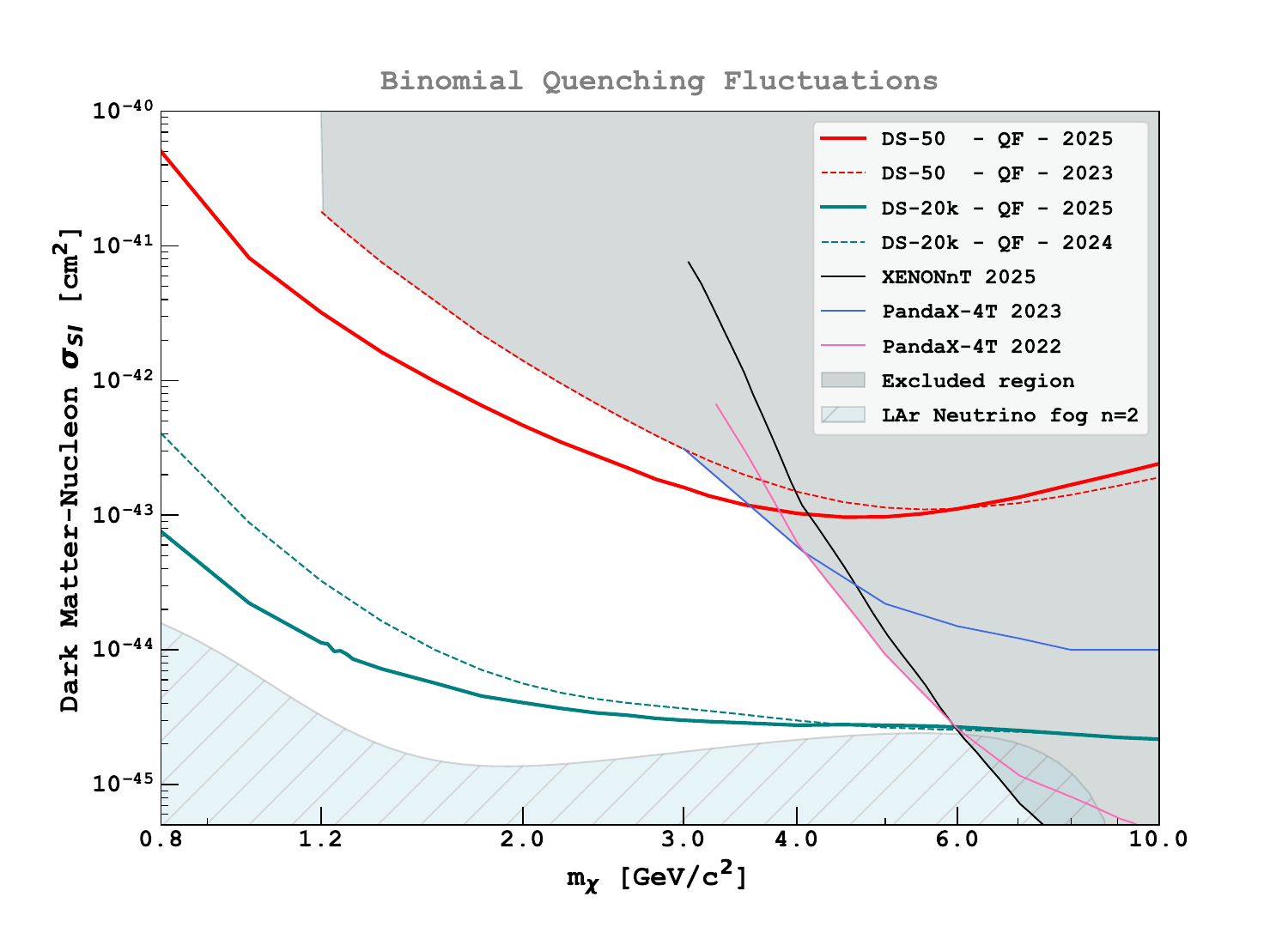}
\caption{
Updated 90\% C.L.\ exclusion limits for DarkSide-50 (solid red) and projected sensitivities for DarkSide-20k (solid teal), obtained with the improved LAr ionization model based on the Lenz--Jensen screening function. Left: no-quenching-fluctuation (NQ) scenario. Right: quenching-fluctuation (QF) scenario. Previous ZBL-based limits and leading results from other experiments are shown for comparison~\cite{DarkSide-50:2025lns}.
}
\label{fig:limits}
\end{figure}

For DarkSide-50, which employs a 4~$e^{-}$ analysis threshold and an effective exposure of 12~ton$\times$day, the updated ionization response yields a significantly larger predicted signal rate at recoil energies below 5~keV. As a result, the 90\% C.L.\ exclusion limit improves by up to a factor of 5 at a WIMP mass of 1.2~GeV/$c^{2}$ under the quenching-fluctuation (QF) hypothesis, and by a factor of 2--3 under the no-fluctuation (NQ) assumption, as shown in Fig.~\ref{fig:limits}. The experiment now sets the strongest published constraints in the 1--3~GeV/$c^{2}$ mass range.

We also updated the sensitivity projections for DarkSide-20k, also shown in Fig.~\ref{fig:limits}, assuming a 2~$e^{-}$ threshold and a 342~ton$\times$year exposure, using the same analysis configuration as in ref.~\cite{DarkSide-20k:2024yfq}. In this regime, the enhanced $Q_y$ prediction substantially increases the probability of detecting the few-electron signals induced by sub-5~keV recoils. For a 1.2~GeV/$c^{2}$ WIMP, the improvement reaches nearly one order of magnitude in the NQ scenario and a factor of 3 in the QF case. The updated projections confirm that DarkSide-20k will achieve world-leading sensitivity to low-mass dark matter.

Overall, the revised LAr response model strengthens both the reinterpretation of existing DarkSide-50 data and the discovery potential of DarkSide-20k, highlighting the importance of a precise understanding of nuclear-recoil ionization at the few-keV scale.

\begin{acknowledgments}


The author acknowledges support from the French Institut National de Physique Nucléaire et de Physique des Particules (IN2P3) and from the French National Research Agency (ANR) through Grants ANR-22-CE31-0021 (X-ArT) and ANR-23-CE31-0015 (FIDAR).

\end{acknowledgments}

\bibliographystyle{unsrtnat} %
\bibliography{biblio}
\end{document}